**An Efficient and Balanced Platform for Data-Parallel Subsampling Workloads**

**THESIS**

Presented in Partial Fulfillment of the Requirements for the Degree Master of Science in the Graduate School of The Ohio State University

By

Satya Sundeep Kambhampati

Graduate Program in Computer Science and Engineering

The Ohio State University

2014

Master's Examination Committee:

Dr. Christopher Stewart, Advisor

Dr. Rajiv Ramnath, Co-advisor

# Abstract

With the advent of internet services, data started growing faster than it can be processed. To personalize user experience, this enormous data has to be processed in real time, in interactive fashion. In order to achieve faster data processing often a statistical method called subsampling is adopted and such workloads are called subsampling workloads. Subsampling workloads compute statistics from a set of observed samples using a random subset of sample data (i.e., a subsample). Data-parallel platforms group these samples into tasks; each task subsamples its data in parallel.

Current, state-of-the-art platforms such as Hadoop are built for large tasks that run for long periods of time, but applications with smaller average task sizes suffer large overheads on these platforms. Tasks in subsampling workloads are sized to minimize the number of overall cache misses, and these tasks can complete in seconds. This technique can reduce the overall length of a map-reduce job, but only when the savings from the cache miss rate reduction are not eclipsed by the platform overhead of task creation and data distribution.

In this thesis, we propose a data-parallel platform with an efficient data distribution component that breaks data-parallel subsampling workloads into compute clusters with tiny tasks. Each tiny task completes in few hundreds of milliseconds to seconds. Tiny tasks reduce processor cache misses caused by random subsampling, which speeds up per-task running time. However, they cause significant scheduling overheads and data distribution challenges. We propose a task knee-pointing algorithm and a dynamic scheduler that schedules the tasks to worker nodes based on the availability and response times of the data nodes. Since we know the task size and the number worker nodes prior to execution, we decide a few initial data nodes that all worker nodes access. Data is fully replicated across these nodes. Based on the response times from the initial set of data nodes, we estimate the cache interference between task execution and data fetch cycles; the replication factor (number of data nodes) is varied accordingly to meet the SLOs of tiny tasks.

In this document, we discuss the challenges of our proposal and propose a task execution framework that can support tiny tasks with an efficient data distribution platform. We compare our framework against various configurations of BashReduce and Hadoop. A detailed discussion of tiny task approach on two workloads, EAGLET and Netflix movie rating is presented. We also benchmark our framework against similar platforms such as Spark.

**Dedication**

I dedicate this work to my bother-in-law Dr. Prasad Calyam, my parents Mrs. Bhavani Kambhampati and Mr. Sarma Kambhampati, my sisters Mrs. Sudha Kambhampati and Mrs.Aruna Kambhampati and my brother-in-law Mr. Pavan Gorti

iv

# Acknowledgments

This thesis would not have been possible with support and encouragement of many individuals. I would like to thank everyone who helped me in master's journey.

I would like to thank my advisor Dr. Christopher Stewart for his constant support, motivation and insights. Without his constant encouragement and guidance this work would not have been possible.

I sincerely thank my co-advisor Dr. Rajiv Ramnath who guided me throughout my research work. His suggestions were invaluable and directed me towards right goals.

Special thanks to Dr. William Stewart and Nationwide Children's Hospital for providing data for my research.

Finally, I would like to thank my friends Vaibhav Krishna, Sudheer Tumu, Prasad Bhandari, Gautham Kollu, Pallavi Chalasani and Akanksha Lonhari whose continuous encouragement helped pursue my master's dream.
.

**Vita**

2008 to 2012 …………………… B.Tech. Computer Science and Engineering,

VIT University, India

2012 to present ………………… Graduate Research Associate, Department of

Computer Science and Engineering,

The Ohio State University

**Publications**

Sundeep Kambhampati, Jaimie Kelley, William C.L. Stewart, Christopher Stewart, and Rajiv Ramnath, Managing Tiny Tasks for Data-Parallel, Subsampling Workloads. In IEEE International Conference on Cloud Engineering Boston, MA, 2014

**Fields of Study**

Major Field: Computer Science and Engineering

**Table of Contents**

# List of Tables

# List of Figures

# Chapter 1:  Introduction

Not every data-parallel platform suits every application. With the increasing need for data-parallel interactive jobs like statistical subsampling workloads, there is a need for building an efficient and balanced data-parallel platform. Statistical subsampling workloads randomly sample a large set of data and perform statistical operations such as mean, mode, etc. These kinds of workloads are increasingly becoming popular with the advent of internet services, mobile devices and sensors. In order to personalize user experience for various services in real time, interactive data-analytics are to be performed [47], [48]. These services generate vast amounts of data. In fact, in the last two years 90% of all the data in the world has been produced [7], [17]. For example, genome sequencing which is a classic subsampling problem has become 50% cheaper in the last 5 years [25]. To express it in in terms of data size, it can generate up to an Exabyte of data. For a real time analysis on these vast amounts of data every point in the data set can't be evaluated.

Real time analysis of such vast amounts of data is often required. For example, consider a scenario where a serious epidemic disease breaks out and we need to find a disease causing genome in the genome sequence. For that several families' genetic data needs to be sampled and AT/CG base pair has to be studied. This should happen in real time. Another scenario, for a personalized shopping experience, vast amounts of user data needs to be studied in real time and this often needs to be interactive.  In both the cases data needs to be subsampled and studied. Processing these kinds of subsampling workloads is challenging. By subsampling a dataset, we have to trade speed for accuracy. Subsampling a dataset makes the computations interactive and can operate on big-data workloads. However, by subsampling a dataset we incur some statistical error.

Subsampling workloads is well suited in the map-reduce model. Data set can be partitioned on to several nodes and each map task subsamples on these partitions and produces intermediate results. Reduce tasks combine these intermediate results. Existing map-reduce frameworks like Hadoop can be adopted [40], [46]. However, map-reduce frameworks such as Hadoop are typically designed for batch processing and don't scale well for subsampling workloads. In typical map-reduce job such as Hadoop or any batch processing frameworks map tasks have very large granularity of data partition. Since subsampling map task randomly partition the data, there will be more L2 cache misses, thereby forcing it to have more memory fetches.

Our key insight is that subsampling workloads benefit from tiny tasks, i.e., map tasks that randomly sample from only a small portion of the sampled data stored on a node. Tiny tasks (tasks that operate on small partition of data) reduce processor cache misses caused by random subsampling, which speeds up per-task running time. These tasks finish in hundreds of milliseconds to seconds. Tiny tasks pose problems at several levels. Our work focuses on four key problems namely, task sizing, scalable task scheduler, launch overheads and scalable distributed file system. In the following paragraphs these problems and proposed solutions are discussed. In Chapter 3 detailed system design addressing these problems will be discussed.

## 1.1 Challenges in processing subsampling workloads

### 1.1.1 Task Sizing

In large tasks, task scheduling overhead is amortized by per-task delays. However, in subsampling workloads large, tasks face large cache miss rates [1]. For tiny tasks, we need a kneepoint where there is a balance between task scheduling overhead and cache-hit rate. We size tasks at the smallest kneepoint on the task size to miss rate curve. The smallest kneepoint is the largest task size before the first increase in the cache-miss rate. We have developed on offline task sizing approach that gets the kneepoint for a dataset.

### 1.1.2 Scalable Task scheduler

To schedule these tasks we need a scalable and dynamic scheduler that schedules hundreds of tiny tasks instead of few large tasks. If we have a completely dynamic scheduler that schedules one task per node at any point of time, then the nodes might have to wait for the next job. A few milliseconds wait time on a millisecond job would be significantly higher. So, we designed a two-step scheduler. In the first step, tasks that are designed based on task sizing approach are randomly assigned to nodes. However, only one task is assigned in the first step. In the second step, we have a feedback loop that analyzes the initial tasks and assigns bunch of task at a time to each node. The dynamic scheduler now queues multiple tasks to a node such that a node need not wait for next task, instead it can quickly fetch from the queue.

### 1.1.3 Launch Overhead

Launch overhead is the time taken for a task to start executing. For example, in a java process JVM start time is a launch overhead. For mitigating task launch overhead we choose BashReduce, a very lightweight implementation of the map reduce paradigm based on running tasks within the Bash shell [10]. These tasks are connected through simple TCP pipes using the nc6 tool. Task-level fault tolerance has never been supported because it adds unnecessary overhead of monitoring a short running task. On failures, entire job is restarted, rather than a task.

### 1.1.4 Scalable Distributed File System

Time needed to read input data should not be a significant factor compared to task durations. Task should not wait for the data. Since these workloads are commonly I/O bound, we expect task runtime to be dominated by time taken to read input data and also, data is available even before the task starts. Using small data blocks for tiny tasks requires us to move away from traditional distributed file systems like HDFS because, using small data blocks is not a scalable approach for these file systems. Though, HDFS allows tasks to read only part of a block, having multiple tiny tasks that operate on the same file block limits parallelism. So we need a distributed in-memory storage system that would have significantly low fetch time compared to job execution time. Also, we need to pre-fetch the data ahead of the task execution so that task wouldn't wait for the data. We can achieve that by using the task scheduler. Since the tasks are assigned in bunches, we can pre-fetch the data based on scheduling. While a task is being processed, data can be fetched for the tasks in the queue. Since, we know the task size and the number worker nodes prior to execution, we decide a few initial data nodes that all worker nodes access. Data is fully replicated across these nodes. Based on the response times from the initial set of data nodes, we estimate the cache interference between task execution and data fetch cycles; the replication factor (number of data nodes) is varied accordingly to meet the SLOs of tiny tasks.

**1.2 Platforms and Workloads**

To evaluate our tiny task approach on subsampling workloads we setup two map-reduce frameworks and compared our task sizing approach, task scheduling and launch overhead. We setup Hadoop and BashReduce (a lightweight implementation of the map-reduce paradigm [10]). We configured Hadoop with three different settings, vanilla Hadoop, Hadoop without task level monitoring and speculative execution and Hadoop with less HDFS (Hadoop distributed file system) interference. BashReduce has only one configuration. Vanilla Hadoop took approximately 4X longer to start tasks compared to BashReduce. A second version of Hadoop had reduced overheads, and third version of Hadoop achieved very low overheads. Scalable distributed data platform was built using Cassandra. Cassandra is highly scalable and distributed key value store [44].

We set up two subsampling workloads. EAGLET (Efficient Analysis of Genetic Linkage: Testing and Estimation) finds disease genes from subsamples of dense SNP linkage data, within the DNA of sampled families [34]. Our Netflix workloads [43] have movie ratings by users and user-rating patterns are studied across movies and over several years. With low overhead and tiny-task sizing, our BashReduce platform sped up EAGLET and Netflix workloads by 3X and 2.5X compared to vanilla Hadoop. We achieved 25% speedup compared to a lightweight Hadoop setup that had low overhead but no task sizing. Our platform achieved 12X speedup on small input sizes where whole jobs complete within minutes, making our platform attractive for workloads governed by service level objectives [27], [32], [42].

On the EAGLET workload, our platform achieved 117 Mb/s per 12-core node, comparing favorably against competing map-reduce platforms for secondary genetic analysis [30], [31]. Throughput scaled linearly as we allocated additional resources. Our platform also scaled linearly within virtualized environments. In a heterogeneous environment, our platform was limited by the last task to finish its work. For small jobs, throughput degraded proportionately to the slowest task to complete. For larger jobs, however, tiny tasks facilitated workload stealing, erasing slowdowns [2], [39], [41].

**1.3 Contributions**

This thesis focuses on interactive, dataparallel workloads [16], [21], [26], [27], [30], [32]. Map and reduce tasks within these workloads complete quickly, relying on efficient processing and on low scheduling overhead [27].

Key contributions

- We quantified cache miss rate as task size increases and made a case for tiny tasks in subsampling workloads.

- We measured scheduling overheads, both run time costs and startup costs on tiny tasks across two map-reduce frameworks.
- We implemented a task sizing algorithm and developed a dynamic task scheduler within BashReduce scheduler. This reduced runtime overhead.
- We developed a scalable distributed file system on top of Cassandra and tailored it specifically to BashReduce.
- We experimentally validated our improved BashReduce platform, comparing it to vanilla and lightweight Hadoop setups across multiple workloads and diverse clusters.

In the remainder of this thesis, Chapter 2 discusses related work. Chapter 3 describes subsampling, task size, and their effect on cache locality and benchmarks per-task overheads in widely used data-parallel platforms. These overheads led us to integrate task sizing within the BashReduce scheduler. Experiments in chapter 4 show that our improved BashReduce achieves high throughput and responsiveness. Chapter 5 concludes.

## Chapter 2: Related Work

It is challenging to maximize processor utilization, reduce network congestion and increase in-memory data fetch at the same time in parallel. This is especially true for interactive workloads. In this section, we describe recent papers on scheduling algorithms, data storage architectures, modeling approaches, and workload-specific designs. These papers advanced the state of the art for interactive, data-parallel platforms. In comparison, this thesis targets subsampling, data-parallel workloads. We show that task size affects data access times and design a platform and scheduler to support tiny tasks.

With increased use of large clusters, data sharing and task scheduling have become increasingly challenging. Previous works such as Sparrow [26], [27] addressed them. Key idea of these schedulers is reducing the granularity of the task (Splitting larger jobs into small/tiny tasks) and this would improve the utilization of compute cluster resources and reduce waiting time for jobs. According to Sparrow [26], tiny tasks help in mitigating the skew created by the tasks that run longer than other tasks in a job. For example, if there is a long running task and several small jobs are waiting for it to complete and overall execution time increases. Also tiny tasks increase the interaction in batch process. There would frequent small accesses to I/O instead of a long job holding an I/O resource and making others jobs for it to complete [26].

In recent works, to mitigate skews in processing these tiny tasks power of-two load balancing [23] have been proposed. Skews/Outliers are either caused by uneven data allocation or by high computational complexity in tasks. If it due to data allocation, then splitting the tasks into smaller would evenly spread the recourses among all the cluster recourses and thereby improving resource utilization as well as reducing the overall execution time. If is due to computational complexity, then these tasks can be allocated a machine/node that has fewer tasks. Replication for predictability works [3], [16], [32], [45] sends requests to multiple nodes and takes the first response, masking transient delays.

For a task to be more interactive, higher resource utilization may not be possible. But by splitting the job into tiny tasks, these tasks will have to wait less time for I/O or any interaction and also guarantees higher resource utilization. However, making a job interactive add network overhead even within local networks. Network and application interactions are opaque. This makes network and application interactions a hard issue to resolve [9]. Using vertex queues, Mizan [18] provides a high throughput scheduler for Pregel workloads and balances network I/O.

For interactive jobs that are I/O bound previous works such as ThemisMR [30] have implemented a 2 I/O property. Minimizing the number of I/O operations is critical to improving performance of interactive jobs. Minimum number of I/O operations required to process a data are two (one for reading, one for writing) defined as 2-I/O property. ThemisMR reads and writes data records to disk exactly twice, which is the minimum amount possible for data sets that cannot fit in memory. [16] extended this approach to networked in-memory storage, attempting to balance network reads relative to processing demand.

Moore's law proves that exploiting parallelism offers diminishing returns for execution time. Using more parallel resources than required, to meet service level objectives do not result in performance improvement. Zhang et al. [42] is a platform for interactive map-reduce. It models execution time for Pig as a function of parallel resources used. These kinds of performance models for resource utilization are useful for making online management decisions [33]. Accurate performance models focusing on intermittent renewable energy for reduce carbon footprint [11] are created in GreenHadoop and GreenSlot [14], [15]. AMAT (average memory access time) is a simple model that makes a strong point: faster storage can significantly decrease execution times. RDD [40], Data Cube [24], Pig [42],FCS [36] and [16] lower execution times by using main memory for storage, rather than disk. However, main memory is volatile and costly. Often, it is paired with disk or SSD in hybrid storage. Tsai et al. [35] provide a framework to compare caching and partitioned hybrid architectures. hStorageDB is one such hybrid system [20].

Graph workloads often run tasks starting from the same vertex multiple times. Weights or edges from the vertex change slightly with each iteration. Results from previous iterations are reused. Data mining and machine learning workloads have similar properties. McSherry et al. [21] propose language support for differential dataflow, a paradigm that allows programmers to specify incremental structure in their programs. RDD [40] users can call functions on cache misses, allowing for certain types of incremental workloads. Waterland et al. [37] cache results for parallel applications transparently within the operating system. Non-determinism presents a challenge for the above approaches. For example, results for our subsampling workloads are not easily cached by input data alone. One solution would cache random-seed keys along with data, but this may disturb the statistical power of subsampling. Other recent work has studied the efficiency of cloud caches, especially for data-parallel workloads [5], [8], [17].

**Chapter 3: System Design**

This chapter describes the system design in details. In section 3.1, subsampling workloads will be introduced in the context of map-reduce model. Detailed analysis of challenges in fitting subsampling workloads in map-reduce model will be described. In section 3.2, tiny task model that addresses the challenges in fitting subsampling workloads in map-reduce model will be discussed. Results showing how subsampling workloads benefit from tiny task model and algorithm for sizing the tasks will be discussed. In section 3.3, two map-reduce frameworks, Hadoop and BashReduce will be studied in tiny task context and a scalable scheduler for BashReduce will be presented. In section 3.4, task level recovery for BashReduce that suits for tiny task will be discussed. In section 3.5, scalable distributed file system built on top of BashReduce will be described.

**3.1 Map-Reduce for subsampling workloads**

Subsampling is a statistical approach that computes means, modes, and percentiles using only randomly selected portions of each data sample. Subsampling workloads can run on data-parallel platforms, e.g., Hadoop, in map-reduce jobs. These platforms scale out by partitioning sampled data across multiple nodes. Each node subsamples within map tasks, producing intermediate results from randomly selected data. Reduce tasks combine these intermediate results. Figure 1 depicts and labels stages for data-parallel subsampling. For subsampling workloads, input data is grouped by a unique identifier or a key (e.g., in a genome data it can be family id of genome). Each subset of data grouped by unique id is called a sample. Sampled subset of data is much smaller than original gathered data. In figure 1, logical size variation between sample and subsample is depicted.

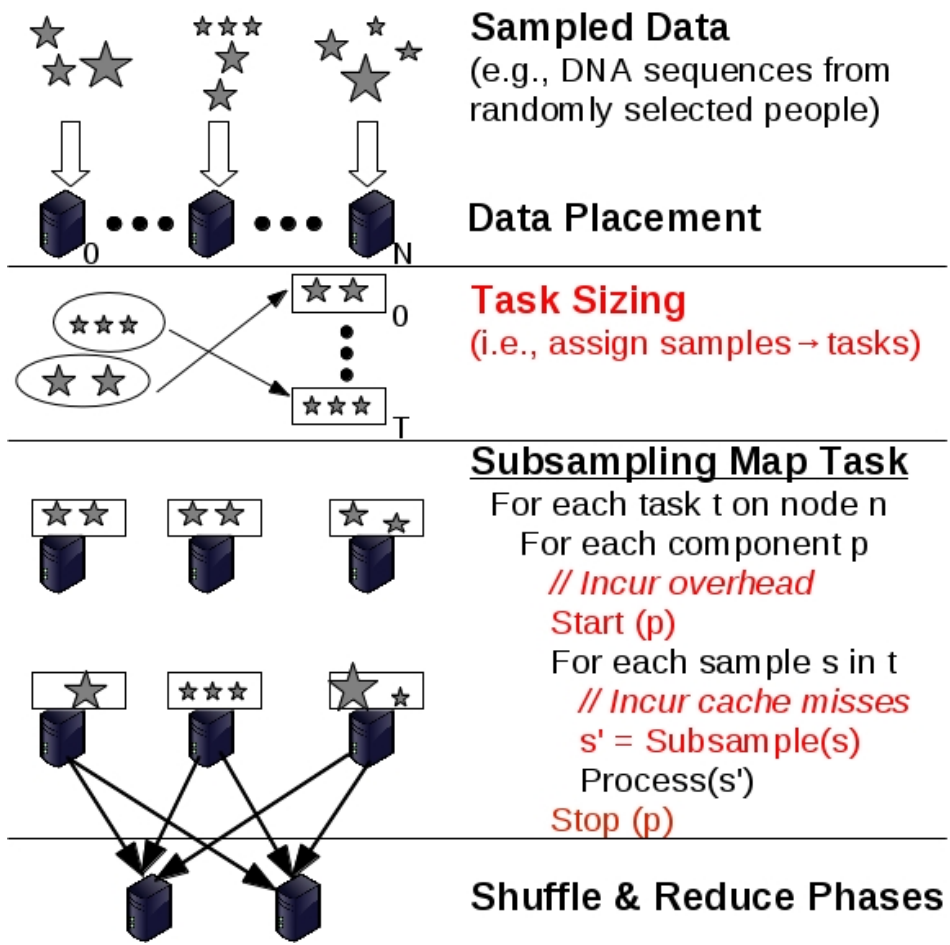

Figure 1: Data flow for subsampling workloads

As the data being processed for these statistical computations is huge data-parallel platforms tends to place data across multiple nodes and compute clusters process this data in parallel by fetching them from multiple nodes. However, when compute clusters access the data from remote node it becomes slower and data fetch time remains as a bottle neck for data processing. Hence, data placement plays a prominent role in performance. Best data placement in terms of computational efficiency is placing all the data on all nodes. This eliminates remote disk access, so that no computational node has to wait for data. However, this is a potential burden on disk space. Considering data to be as big as few hundreds of tera-bytes, placing entire data on each node isn't feasible. However, such full replication is only feasible for small datasets. A more disk efficient approach would be to distribute data across few nodes and replicate it on few nodes. In practice, each sample is stored on only a few nodes and some nodes store more samples than others. Such data skew will cause remote data accesses when nodes with few samples try to steal work from heavily loaded nodes [2]. Previous works [2], [29] focused on handling data skew and load balancing and handling. Our research does not contribute to this part of data placement research.

A task is a collection of software components to process a statistical computation. ($p$ in Figure 1) Number of samples that are processed in a single invocation of software component is called task size. . Number of samples on a node is represented by Sn. Task size of Sn starts each software component in the task once and pipes all results between components. If the task size is equal to or close to Sn, resulting task size is called large task. Advantage with large task is, it starts each software component only once on a single node. It incurs no scheduling delays, minimal launch overhead per node and no intermediate temporary file transfer. However, subsampling workloads present a challenge: subsampling software components have a random data access pattern. This results in poor data locality and cache locality problem. So each computational node has to fetch data from several data nodes. Larger the task size bigger is the data locality problem.

If we consider other extreme of making task size to 1, meaning each software component processes only on sample at any given point of time. Setting task size close to 1 makes a task tiny task. We define tiny tasks as tasks with size close to 1. Tiny tasks have an advantage, they have very high cache locality compared to large tasks. However, they have significant scheduling overhead, have task launch overhead per node and huge intermediate file transfer cost.

This thesis focuses on task sizing for data-parallel subsampling workloads that have a service level objective to complete in minutes to several seconds. Many platforms that support task sizing model store the data in memory. This ensures low data access costs. Sparrow [27], [16], Pig [42], Data Cube [24] and RDD [40] are few examples of platforms supporting task sizing. These workloads may support personalized advertising, sentiment analysis, real-time trace studies and interactive analysis of scientific data [42]. Each task, be it large or small, produces intermediate results. These results are sent to an intermediate shuffler and then sent to a reducer. Reduce phase for interactive workloads are much shorter compared to map phase. Task sizing is efficient for tasks that have relatively shorter reduce phase. If reduce phase is large tasking sizing has low impact [41].

**3.2 Tiny task model for subsampling workloads**

If the location of the data is known prior to processing the data can be pre fetched into processor cache. This increases the cache locality and reduces the data fetch dependency bottle neck in distributed data distribution platforms. This is want generally happens in traditional data-parallel workloads. On the flip side, for non-traditional workloads like subsampling workloads that we are targeting this is not the case. Subsampling tasks decide which data is accessed in runtime. Data access patterns are random and only part of the available data on disk is used. There is no way that a programmer can know which data will be fetched prior to starting a software component. So data can't be pre fetched. Cache misses caused by these random accesses are proportional to task size. As task size grows, more processor cache misses are likely. For large task size incur more random accesses and this translates to higher number of processor cache misses.

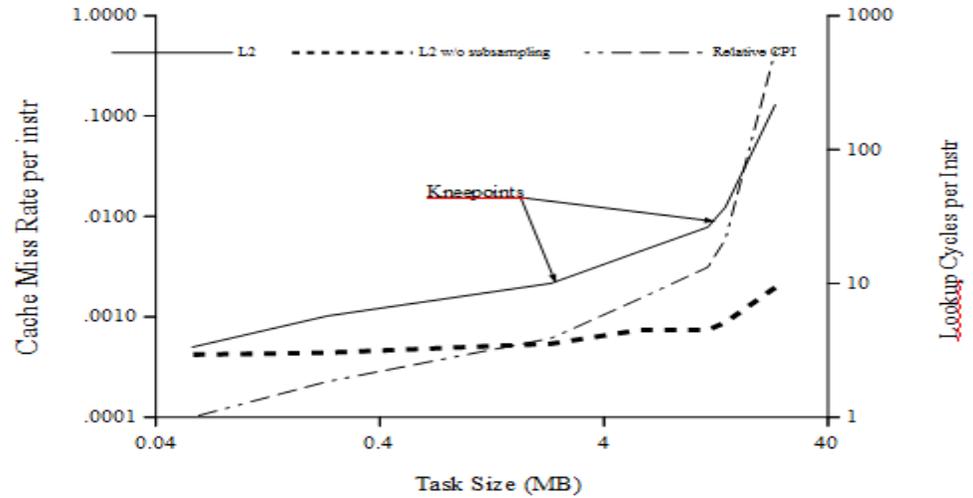

Figure 2: L2 misses per instruction and cycles per instruction across task sizes in EAGLET

Figure 2 makes the case for tiny tasks on the EAGLET subsampling workload. EAGLET (Efficient Analysis of Genetic Linkage: Testing and Estimation) finds genomic sequences correlated with diseases [34]. It is helpful for detecting the disease related genes in human Genome. EAGLET computes average LOD-score curve based on multiple subsamples of dense SNP data. LOD score curves are calculated across multiple subsamples of SNPs, where each subsample is chosen so as to minimize linkage disequilibrium (LD) while simultaneously maximizing the information for linkage). Then, the linkage results from each of the highly informative subsamples are combined over a common grid to form the ALOD.

For this purpose of identifying disease gene Eaglet needs the genes of individuals from at least one generation of family (This data is referred as SNP data). One-generation data consists of father mother and their children (Ideally it would good to have data about 2 children. Data with one child may not contribute to the statistic). As the number of generations of a particular family increases it would contain more information and may take longer processing time as well. Subsampling can be done on per family basis at the least. So, one family would the atomic part for computing the statistic. So the data for a single statistical computation would be would a combination of families ranging from one family to all families in the data set. We need to find the optimal number of families such that computational time is reduced and resources are efficiently used.

We studied linkage data from bi-polar diseases. We initially started with 230MB of real data. This makes 400 samples. Then we scaled it up to few tera-bytes as needed. More detailed experimental setup will be discussed in chapter 4. In practice, Before requesting costly lab analysis on this genome data scientists use EAGLET. First they use EAGLET to determine disease causing step to detect disease genes. Scientists may use EAGLET to test up to 105 genomic sequences for statistical correlations. To allow scientists to interactively refine their hypothesis EAGLET jobs should compute the statistic as quickly as possible.

In Figure 2, the task size presented in MB reflects the number of families included in EAGLET's input list. At runtime, EAGLET randomly selects subsets of SNPs from a family's genome. ALOD scores for the subset of data are computed. This will mark the disease causing genome and computes intermediate results. Reduce phase is fed with these ALOD scores. Reducer combines these ALOD score and generates the LOD score required for scientists. This entire process is of computing LOS scores involves several software components written in several programming languages. MERLIN, Perl, GenLib are key components among several others. To measure processor cache misses while EAGLET ran, we used OProfile [19]. We ran these experiments on an Intel Sandy Bridge processor with 6 dual cores with 1.5MB L2 cache and 15MB L3 cache. We observed that large tasks incurred higher miss rates. A 25MB-sized task saw 35X more L2 cache misses per instruction than a 2.5MB sized task. The EAGLET subsampling component is the source of the increase missed rate. The miss rate was flat among other components.

There are two ways in which random accesses increase the cache miss rate. First, a compulsory cache miss is incurred if the data being accessed is not in processor cache and this is more likely to happen if the data size is huge. Second, in LRU caches it is likely that random access patterns have increased processor cache misses [12]. Stack distance is the number of unique data references between accesses to the same data. Stack distances smaller than the cache size means data accesses will hit in cache. Random accesses (due to subsampling) injected between normal accesses make cache hits less likely. This explains a key property of Figure 2: The miss rate changes at certain key task-size thresholds. After those points, increasing the task size results in random accesses evicting frequently accessed data that normally, i.e., without subsampling, would have hit in cache. We call points where the miss rate increased sharply kneepoints. Kneepoints were at 2.5MB and 11MB.

Additionally, we also captured L3 cache misses. Kneepoint for L3 cache was observed at 11MB. If there is a cache miss, program is forced to fetch data from memory. And memory fetch is 63 times slower than L2 cache fetch on architectures such as Intel Sandy Bridge. Average memory access time (AMAT) per instruction, the time for a lookup in the fastest cache plus the product of the miss rate and the miss penalty, is a well-known model to study the effect of cache misses [28]. The secondary axis on Figure 2 plots the normalized AMAT where the fastest cache looks up results in 1 cycle. We observed over a 1,000X increase in AMAT between the tiniest task and the largest task.

3.2.1 Task sizing algorithm

In the earlier section we made a case for tiny tasks for subsampling workloads. In this subsection we will be describing a task sizing approach for subsampling workloads. Many tiny tasks in a software component translates to more per-task scheduling overhead than few large tasks. Per-task delays such as starting JVM in java process, across many samples, will be amortized in large tasks. However, as shown in earlier section very large tasks face large cache miss rates. We size tasks at the smallest kneepoint on the task size to miss rate curve (i.e., Figure 2). The smallest kneepoint is the largest task size before the first increase in the cache-miss growth rate. Our approach achieves low cache miss rates while amortizing per-task overhead across samples. Our approach involves an offline step and an online step, we created the task size to miss rate curve and found kneepoints. In an online step, we packed subsamples into tasks.

Specifically, Figure 3 outlines our approach. First, during an offline phase, we collect data on the relationship between task size and cache misses. On a benchmarking node, we run Oprofile. We run map tasks in isolation, varying the number of samples in the task's working set. As seen in Figure 2, we plot the aggregate input data size against cache misses per instruction. We modified our platform to group samples into tasks of equal (kneepoint) size before starting map tasks. We place the same number of samples in each task, assuming samples are roughly the same size; in practice, data parallel jobs have large outliers [3], [16]. Our genetic analysis dataset also has outliers, with one sample 15X larger than the mean and a second sample 7X larger than the mean.

The time taken by the offline phase is about 3% of the time taken by the online phase. However, the offline phase is a one-time overhead paid for each new data set. We compared the impact of task sizing on BashReduce's performance. We ran EAGLET on the 72-core Sandy Bridge cluster. EAGLET subsampled data and computed genetic statistics 30 times for each family. Each of these subsamples (i.e., 30 x 400 families) could run in its own map slot. Figure 4 shows throughput relative to 24MB large tasks, i.e., the amount of data partitioned to each map slot in the cluster (Sn). Our results include the delay for determining the kneepoint offline.

```
                    Offline: Determine Kneepoint
public static int kneepoint(int maxSampleNum) {
  float[2] taskSizes = new float[];
  float[2] missRates = new float[];

  //Pick random samples for study
  float[] samples = RandomArray(1, maxSampleNum);
  List workingSet = new List();
  workingSet.add(samples[0]);

  // Run the tiniest task and collect misses
  results = ExecTask(workingSet);
  misses[0] = results.cacheMisses();
  taskSizes[0] = results.inputSize();

  int growthRate = 0, i = 0;
  float MAX_RATE = -1;
  // Run tests at each size, compare miss rates
  while ((growthRate <= MAX_RATE) ||
      (MAX_RATE == -1)) {
    workingSet.add(samples[i]);
    results = ExecTask(workingSet);
    missRates[1] = results.cacheMisses();
    taskSizes[1] = results.inputSize();
    growthRate = ((missRates[1] – missRates[0]))
                  /((taskSizes[1] – taskSizes[0]));
    //bookeeping
    if (MAX_RATE == -1) MAX_RATE = growthRate;
    missRates[0] = missRates[1];
    taskSizes[0] = taskSizes[1];
    i++;
  }
  return (taskSize(i-1));
}
```

```
                    Runtime Scheduler: Task Sizing
public void sizing(int kneepoint,
            InputStream dataset) {
  // determine size in terms of # samples
  float AVG_SAMPLE_SIZE = K;

  int size = kneepoint / AVG_SAMPLE_SIZE;

  //Split dataset into tasks
  InputStream[] tasks;
  tasks = splitInputStream(dataset, size);
  for(InputStream task: tasks){
     addToMapJobList(task);
  }
  // start Bash Reduce
  StartBashReduce();
}
```

Figure 3: Task sizing algorithm

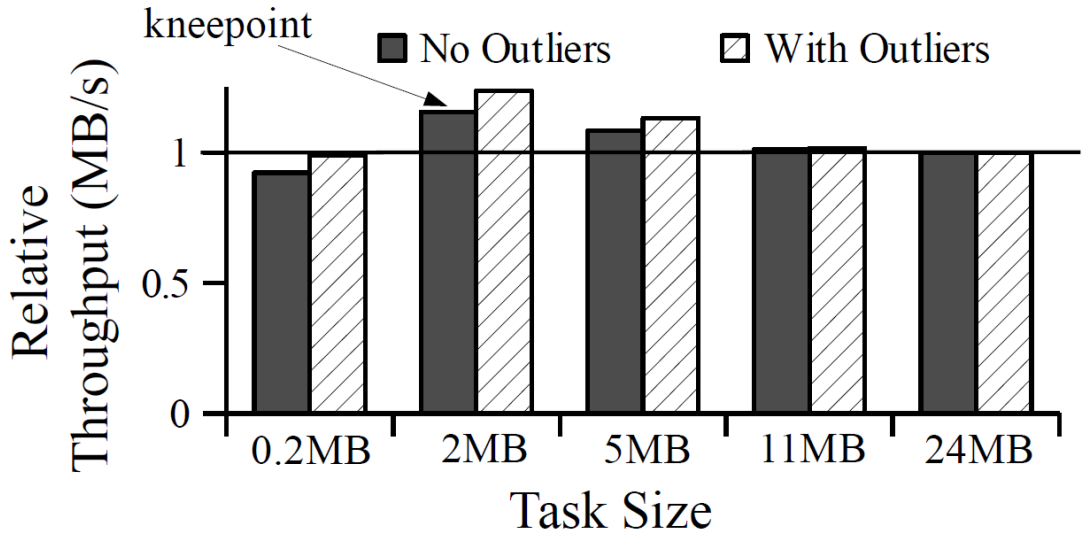

Figure 4: Impact of our kneepoint algorithm on runtime

First, we removed outlier samples from our dataset (shown as no outliers in Figure 4). Outlier samples run 50X longer compared to the mean run time, or longer. We observed that our kneepoint approach achieved 15% speedup compared to the baseline created by the 24MB large task approach. Further, the tiniest task approach caused 8% slowdown. When we included the outlier samples, we observed that our approach increased throughput by 23%. This is because the outlier tasks increased the cache miss rate within their task groups by pushing valuable data out of the cache. Tiny tasks were more helpful under the heterogeneous workload. The absolute running time with heterogeneous tasks under the tiniest task approach was 791 seconds with outliers, and 322 seconds without the outliers. Outliers themselves caused a 2.4X slow down [3], [32].

Our task sizing approach had a larger impact with outliers but did not overcome the slowdown caused by outliers.

The kneepoints identified by our offline analysis are contingent on hardware and workload. The task size to miss rate curve should be recomputed if processor cache sizes or data access patterns change. Our ongoing work attempts to identify a cross-platform heuristic to identify kneepoints, especially for cloud platforms where processor cache sizes are not known. Our experiments in the next section show that kneepoint selection is insensitive to small errors.

**3.3 Job-level Recovery**

Tiny tasks have fewer cache misses per instruction than large tasks. However, data-parallel platforms configured to use tiny tasks will start and stop software components more often than platforms configured to use large tasks. The time taken to schedule software components, called scheduling overhead, may exceed the time saved by improved cache locality.

Hadoop monitors each task's execution for potential node or disk failures. On failure, tasks are restarted with different resources. The monitoring and data replication required for such task-level recovery are major sources of scheduling overhead. Job-level recovery, in which a node or disk failure would restart the whole job, can lower scheduling overhead [30]. In this section, we first make a case for job-level recovery in interactive data-parallel workloads. Then, we quantify scheduling overhead in data parallel platforms, comparing a vanilla Hadoop setup, lightweight Hadoop setups, and a clean-slate platform. We reduce scheduling overhead by moving toward job-level recovery.

Hadoop was designed to process multiple petabytes spread over $10^4$ - $10^5$ nodes [38], taking hours or days to complete a map-reduce job. During the course of a job execution, multiple disks and nodes were likely to fail. If each failure restarted the entire workload, the job would never complete on Hadoop, making the decision for task-level recovery on Hadoop simple.

We revisit task-level recovery here in the context of interactive, subsampling workloads that run for minutes. The shorter time frame makes it $10^3$ – $10^4$ times less likely that a failure will occur in the midst of a job execution. Further, these workloads use fewer nodes because 1) data stored in main memory is costly [16], [27], [40] and 2) their goal is often to compute results from iterative or incremental changes [21], [22].

Mechanisms for task-level recovery, e.g., monitoring and data replication, increase a workload's running time. Let $cost_{tl}$ be the slowdown factor. On failures, only tasks are restarted, rather than entire jobs. On each failure, task-level recovery saves the difference between the expected job and task running times. Our key insight is that task-level fault tolerance only makes sense if 1) hardware failures occur faster than jobs complete, meaning every job is likely to see a failure or 2) rerunning entire jobs would slow down running time by more than $cost_{tl}$. For short, interactive workloads, the latter concern is most important.

Let mttf represent the mean time to a node or disk failure. Also, let _ P(w) reflect of service level objective (SLO) for the workload [42]—i.e., the worst case running time.

We expect at most (fw = N __ P(w) mttf __) failures during an execution. Here, _
captures correlated, heavy-tail failures that occur within the SLO window. We now
compute fw for typical subsampling workloads. We set _ P(w) = 10 minutes and _ = 1:5.
Taking guidance from recent work [16], [27], [41], we set N = 100. We set mttf = 4:3
months from [13], [30]. Under these settings, fw = 0:0078, meaning that monitoring
overhead would have to fall below 1% to justify task-level recovery. Next, we quantify
actual overheads observed in Hadoop.

| Codename | Core | Task-level Failures | Full Dist. File Sys. | Java |
|---|---|---|---|---|
| Vanilla Hadoop | Hadoop | Yes | Yes | Yes |
| Job-level Hadoop | Hadoop | No | Yes | Yes |
| Lite Hadoop | Hadoop | No | No | Yes |
| BashReduce [10] | Unix Utilities | No | No | No |

Table 1: Comparison chart of platforms

**3.4 Platform Selection**

We measured scheduling overhead for the platforms shown in Table 1. Here, we describe the salient features of each platform. More details are can be found in Section IV.

Hadoop was an obvious choice to benchmark, as it is widely used in practice for map reduce workloads. Vanilla Hadoop used default monitoring and HDFS policies. Each task reports its progress to a central service that exposes an HTTP front end. Also, tasks use HDFS instead of the local Linux file system. In the job-level Hadoop setup, we disabled the central monitoring service. In the lite Hadoop setup, we modified EAGLET so that map tasks created no intermediate HDFS files, avoiding replication costs. This new version of EAGLET performed calculations based on a static, globally distributed file rather a dynamic file. We also disabled the central monitoring service in lite Hadoop. Note, lite Hadoop is shown for benchmarking only—its results are incorrect.

The BashReduce platform takes a clean-slate approach [10]. It is a very lightweight implementation of the map reduce paradigm based on running tasks within the Bash shell. These tasks are connected through simple TCP pipes using the nc6 tool. Task-level fault tolerance has never been supported in BashReduce. BashReduce also elides a global distributed file system (HDFS). Managers partition data and tasks access only the local file system.

We quantified two types of scheduling overhead. Startup time captures delays that happen only once for each workload. These delays include TCP handshakes for outstanding connections and data staging. Runtime overhead captures delays incurred as a task runs. Specifically, runtime overhead is the difference in running time between running software components directly on Linux and running them on one of the platforms in Table 1. We ran these experiments on a 72-core cluster consisting of 6 dual-core Intel Sandy Bridge processors. Each core served as a map slot. Task size was fixed at 1 sample.

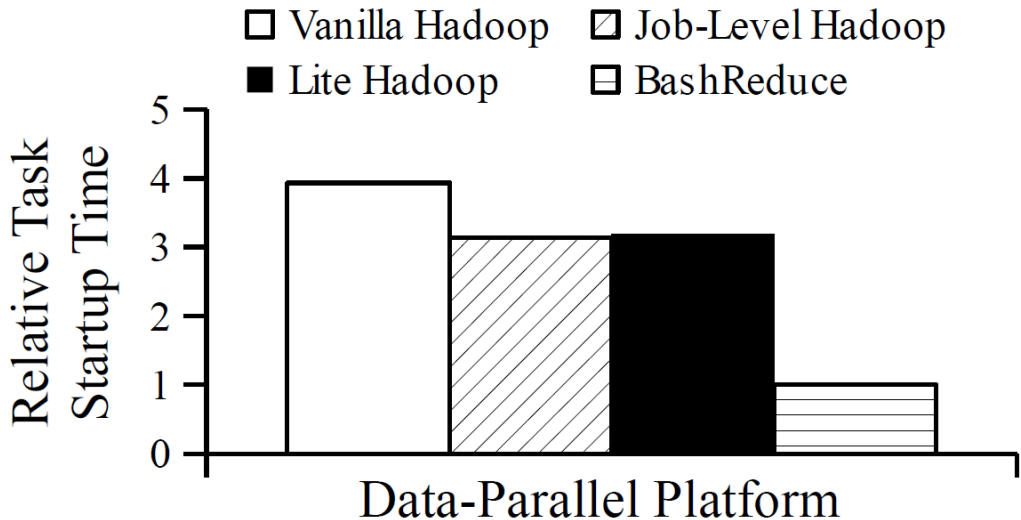

Figure 5: Runtime overhead of each platform relative to native Linux

We measured startup time by running a hello-world job where tasks equaled map slots. Each task was identical and completed within milliseconds (less than 0.01% of the job's running time on Hadoop). Figure 5 shows the time taken to complete this job. Times are normalized to the overhead of BashReduce. Task monitoring overhead increased Hadoop's startup costs by 21%, about 52 seconds. Task-level failures would have to recover hundreds of sub-second subsampling tasks to justify this large overhead. Using formulas from the previous section, clusters smaller than 30K nodes do not justify 21% overhead. BashReduce could start jobs almost 4X faster than vanilla Hadoop.

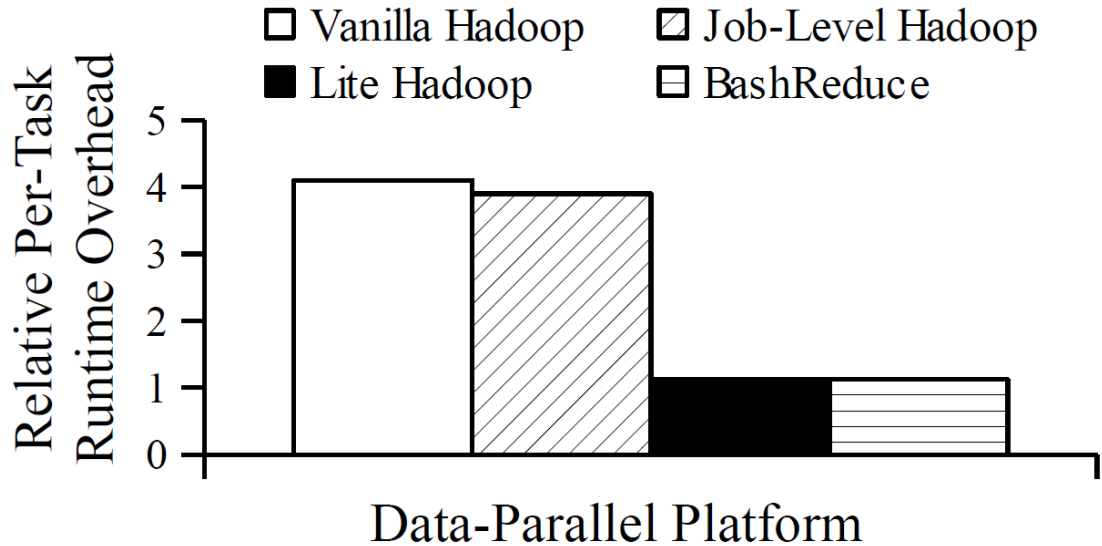

Figure 6: Runtime overhead of each platform relative to native Linux

Figure 6 compares the relative per-task runtime overhead of each platform. For this test, we ran an EAGLET subsampling workload comprised of 4K tasks and measured the total running time. Then we subtracted the startup time and divided by 4K. The result is shown relative to the running time on Linux without a platform. Failure monitoring caused a 20% degradation per task. However, the largest runtime gain came from bypassing HDFS on short-lived temporary files. Indeed, the experiment on Linux without a platform achieved runtime overhead almost equal to BashReduce's overhead. BashReduce still incurred 12% overhead due to scheduling the subsampling map tasks on the cluster. In practice, this overhead would accumulate for tiny tasks. In the next section, we address this overhead by looking for relatively large tiny tasks.

## 3.5 Scalable distributed file system

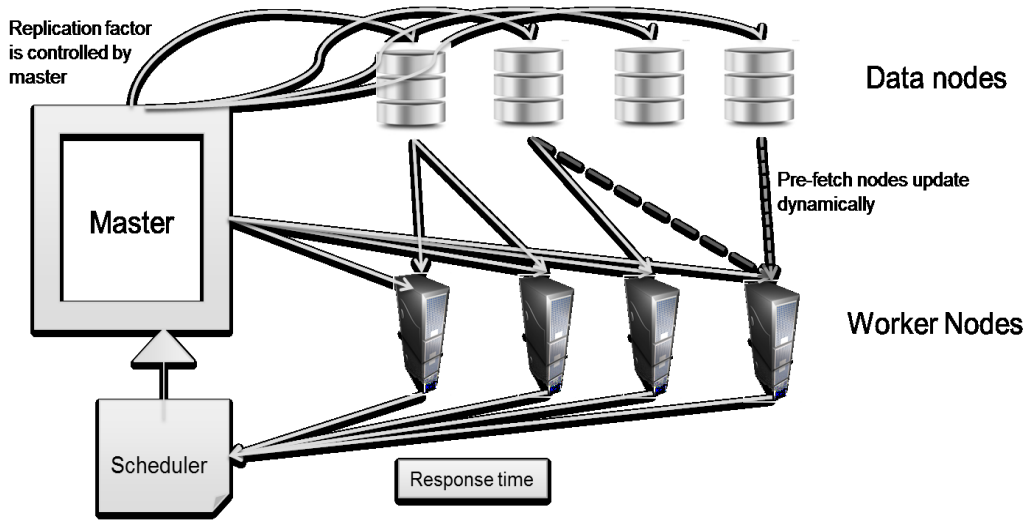

Figure 7: Dynamic Scheduling and Data Scheduling model

For the tiny task model to work data fetch time should not be a bottleneck for data processing. In Vanilla Hadoop, Hadoop Distributed File System (HDFS) is used. HDFS stores the data files in blocks of data. In BashReduce, the underlying file system is used. Both in Hadoop and BashReduce by default full replication is expected. However, this isn't quiet scalable approach. For a TaraByte of data to be available on all the worker nodes would be a disk heavy approach. Using small data blocks for tiny tasks requires us to move away from traditional distributed file systems like HDFS because using small data blocks is not a scalable approach for these file systems. Though, HDFS allows tasks to read only part of a block, having multiple tiny tasks that operate on the same file block limits parallelism.

Naïve approach is to use in-memory key-value store. This partly ensures low fetch time compared to job execution time. If data is pre-fetched ahead of the task execution so that task wouldn't wait for the data. However, lot of pre-fetching would limit the dynamic scheduling of tasks. We can achieve this by using the task scheduler. Since, the tasks are assigned in groups, we pre-fetch the data based on scheduler. While a task is being processed, data required for the next k tasks are pre-fetched fetched. K is decided dynamically from the average data fetch time and average task execution time.

Other key feature is to limit the replication factor. Since, we know the task size and the number worker nodes prior to execution, we decide a few initial data nodes that all worker nodes access. Data is fully replicated across these nodes. A data modelling engine collects the data fetch time from each node. Also, a feedback loop from task scheduler gives the execution time of tasks. Based on the response times from the initial set of data nodes, we estimate the cache interference between task execution and data fetch cycles; the replication factor (number of data nodes) is varied accordingly to meet the SLOs of tiny tasks.

## Chapter 4: Experimental Setup and Results

In Chapter 1, a brief introduction to the research work being presented in this thesis was discussed. In chapter 2, related works that support the work presented was discussed in detail. In chapter 3, detailed system design was presented. In this chapter, experimental test bed of the system presented in earlier chapter will be presented. Various platforms used and compared against will be introduced, followed by hardware details. Then a detailed discussion on various experiments performed on various platforms will be presented. Data showing the performance improvement with tiny task model, tradeoffs between different map-reduce platforms, speedups with respect to task sizing on different hardware and different subsampling workloads will be presented.

**4.1 Experimental Setup**

4.1.1 Workloads

We set up two subsampling workloads, EAGLET and Netflix workloads. EAGLET is a legacy statistical application. Netflix workload was movie average ratings and other similar statistics on Netflix movie rating data.

4.1.1.1 EAGLET

EAGLET [34] (Efficient Analysis of Genetic Linkage: Testing and Estimation). It is helpful for detecting the disease related genes in human Genome. It is open-source software. We initially started with bi-polar human genome data of 400 families constituting approximately 4000 individuals. These individuals volunteered for a research on bi-polar diseases. In this workload, genome sequence from single family (includes all the individuals from a family) is a single sample or the smallest task. The workload recomputed analysis that unveiled well-known linkages [4]. In total, the original data from 400 families is approximately 230 MB. It is a common practice for statisticians in genetic analysis to run a statistical computation for 30 to 50 times for increasing the confidence in computation. So 30 times each sample makes the data set to 6.9GB. As we scaled our experiments we simulated data from the original computation. This data is statistically similar to original data. However, it does not contribute much to finding the disease causing genome. We used the simulated data to scalability of our approach. Our largest test has 684K families and the entire job for 30 samples was a 1 TB. The distribution of family sizes (and hence sample sizes) was heavy tailed. Outliers were preserved in our synthetic data.

4.1.1.2 Netflix

Second workload we setup was based on Netflix movie rating. This data was obtained from [43], [41]. Unlike EAGLET workload this isn't a legacy application. Here, each sample represents a movie that Netflix streamed to its users. The data within each sample are tuples composed of the date, user id, and the user's rating of the movie. Our workload subsampled ratings for each movie to estimate typical user ratings by month. Data size was 2 GB with 118 KB per movie. By subsampling, we found the user ratings faster than exhaustive calculation would have [41] but we also allowed errors to occur. We classified two types of Netflix workloads: High confidence and low confidence. The high confidence workload estimates average user ratings with a 98% confidence interval, choosing less speedup and more accuracy. The low confidence workload estimates use two orders of magnitude fewer ratings, accepting more error for speedup.

4.1.2 Task Sizing

Our EAGLET and Netflix workloads differed in terms of software complexity. EAGLET used multiple (> 5) open-source software packages that spanned three programming languages. Our Netflix workloads used only Bash scripts. We hypothesized that EAGLET was more likely to suffer from tiny-task scheduling overhead.

Both workloads used a pointer to a file containing the actual input data. If the file was large and contained many samples, the task operating on the file was large. If the file was small and contained few samples, the resulting task was tiny. Precisely, we define large tasks as jobs that consist of all of the samples partitioned to a node (i.e., Sn samples in 1 file). The tiniest tasks have Sn files that are piped one-by-one into the respective programs.

4.1.3 Platforms

We compared the following platforms.

1. BashReduce with Task Sizing (BTS): We set up BashReduce [10] with netcat (nc6) for inter-node communication via pipes. BashReduce centralizes scheduling and shuffling stages on a single master node. In our configuration, the master node also decides on task sizes by creating input files locally and distributing them to all other worker nodes. The master node includes the offline script described in Figure 5. Unless otherwise mentioned, BTS sets task size to 2.5 MB for EAGLET and 1 MB for Netflix. If any master or worker node fails, the entire BashReduce job is restarted.

2. BashReduce with Large Tasks (BLT): In this configuration, the master node referred to all samples on a node within a single file.

3. BashReduce with Tiniest Tasks (BTT): In this configuration, the master node referred to only 1 sample in each of Sn input files.

4. Vanilla Hadoop (VH): We compared other platforms against Hadoop, a widely used platform for data analysis. Our default configurations uses an HDFS replication factor of $N_2$ to reduce data migration traffic. A large replication factor is a sensible optimization for interactive workloads that use relatively small datasets. Each node is configured to have as many map slots as cores.

5. Job-Level Hadoop (JLH) disables TaskTracker, the feature responsible for task level recovery. Also, speculative execution is disabled. These optimizations make Hadoop more suitable for our interactive workloads by reducing task startup and runtime overheads.

6. Lite Hadoop (LH): This benchmark produces incorrect results but achieves very low overhead on the Hadoop platform. We use it to benchmark overhead from Java Runtime and to understand the potential for revised subsampling-aware Hadoop. We changed EAGLET so that it fixes intermediate files used to pass data between software components. The subsampling portion of EAGLET was unaffected. We set the replication factor to N on the intermediate files, ensuring no HDFS data transfer would slow down the platform.

4.1.3 Hardware

In table 2, different hardware used in these experiments is shown. Processors include AMD and Intel brands that vary by cache size, memory capacity, and processing speed. We restricted the amount of hardware being used. We focused on the how performance varies with limited hardware. This is often the case with real time systems. Resources are limited. Ideally, maximum performance is expected out f limited resources.

|  | Type I | Type II | Type III |
| --- | --- | --- | --- |
| Processor | Xeon | Xeon | Opteron |
| Cores per Node | 12 | 12 | 32 |
| Processing Speed | 2.0G | 2.3G | 2.3G |
| L2 Cache | 15MB | 15MB | 32MB |
| Memory | 32GB | 32GB | 64GB |
| Virtualized | No | No | Yes |

Table 2: Types of hardware

**4.2 Experimental Results**

Experimental section is organized as follows. First, results on task sizing approach on various workloads are presented. Various configurations of BashReduce are evaluated on both workloads. Kneepointing algorithm on workloads will be shown. Second, both map-reduce platforms Hadoop and BashReduce are compared. For this three different Hadoop configurations and three BashReduce configurations as described in earlier subsection are used. Third, elasticity if task sizing approach is evaluated by scaling the number of cores from 12 to 72. Finally, various heterogeneous platforms shown in table 2 are evaluated. Also, they are evaluated in terms of virtualization.

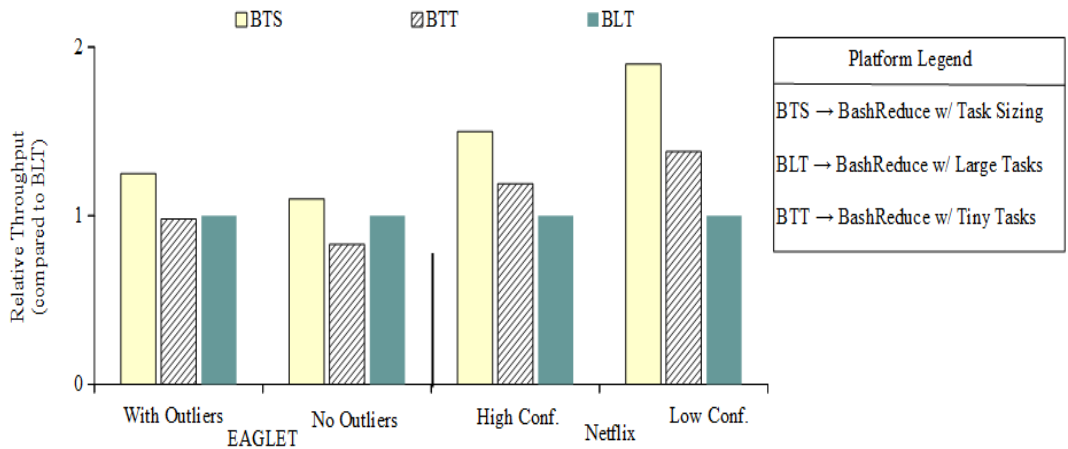

Figure 8: Impact of kneepoint algorithm on runtime

4.2.1 Task sizing on workloads

Figure 8 compares the BashReduce setups. For this test, we used 6 nodes of hardware type 1 (See Table 2). In total, the tests ran on 72 cores. These tests used only the original data from the Bi-Polar study and movie ratings. We observed that BTS achieved throughput 10–90% higher than BLT and 26–32% higher than BTS. Because the Netflix sampling workload uses fewer software components than EAGLET, it was able to better exploit cache locality, resulting in favorable BTT results. In contrast, EAGLET suffered additional per-task runtime overhead from starting many software components on tiny tasks. BTS balances these issues, typically outperforming its closest competitor by 17%.

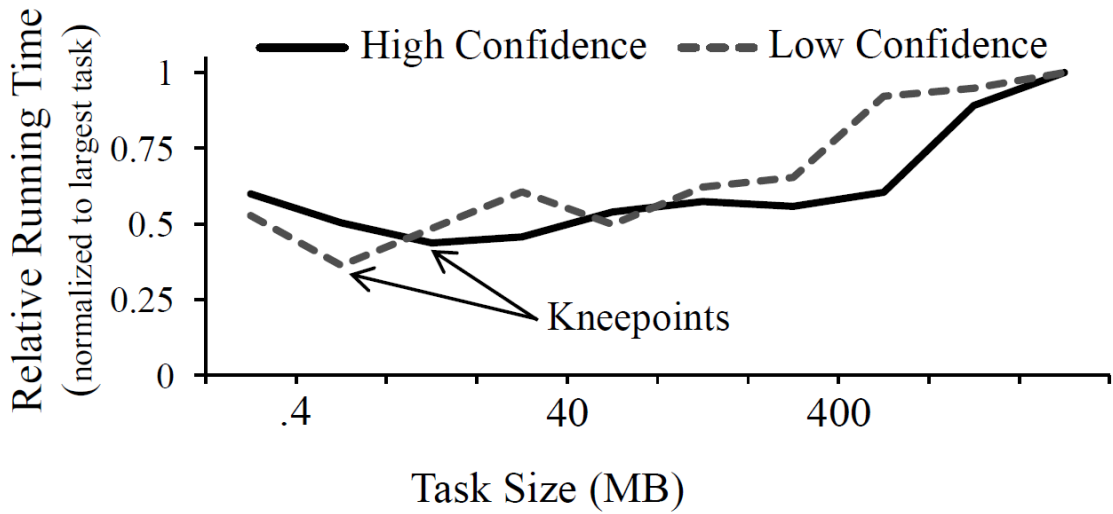

Figure 9: Kneepoints in the Netflix subsampling workloads on BashReduce

Figure 9 shows that kneepoints occurred for the Netflix workloads as well. Results shown were run on top of BashReduce. However, the kneepoints occurred at different places for the high and low confidence workloads despite subsampling the same data. We expected this result because cache locality patterns varied depending on the confidence level desired. Our offline approach can find a different kneepoint depending on the workload, provided the data is available. For results presented in this section, we used only 1 kneepoint (1 MB) for both Netflix workloads. Results with high confidence workload in Figure 7 show that exact kneepoints are not needed to improve throughput relative to BLT and BTT. To quantify how robust our approach is, we created five Netflix workloads that varied according to their output confidence level. Among the five workloads, the 1 MB task size ranked in the top 2 task sizes (in terms of throughput) three times. In the cases where it was not the best performing task size, it was within 10% of the best. Further, the 1 MB task size setting outperformed large and tiniest task settings in all 5 workloads.

4.2.2 BTS versus Hadoop:

Hadoop is a widely used platform for data processing. However, it is not designed for short, interactive jobs [38]. We compared the throughput of BTS to three Hadoop setups across different job sizes. For these tests, we ran the EAGLET subsampling workload on type 2 hardware, varying job size. We changed the job size by adding synthetic families to the Bi-Polar data.

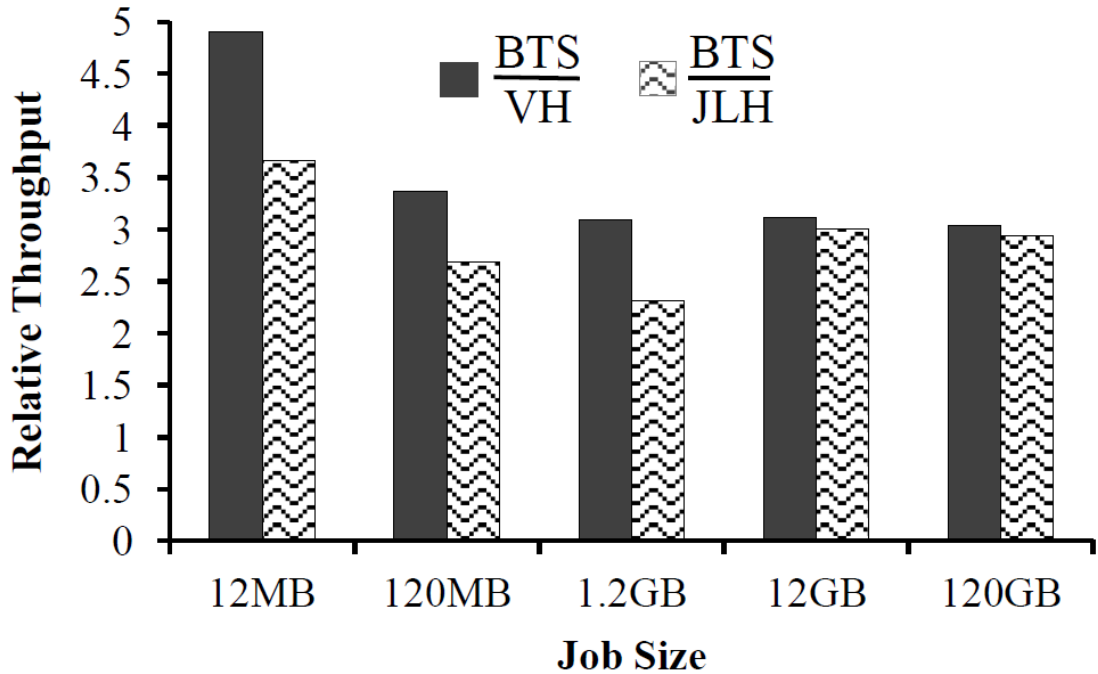

Figure 10: Comparison of BTS to VH and JLH

Figure 10 shows that BTS sped up VH by almost 5X on jobs with a 12 MB task size. For reference, we found that a 12 MB job can test a genetic hypothesis on 40 families with 15 subsamples per family. As the job size increased, BTS offered less speedup because VH was able to amortize its startup costs. We recall here that JLH had lower startup costs and runtime overhead compared to VH. JLH performs better on short jobs, but BTS still offered 3.7X speedup.

Along with tracking task-level failures, the Hadoop platform monitors CPU utilization, I/O efficiency and other system metrics. The metrics are queried frequently to produce user friendly web displays about the state of the system. We added system level monitoring into BTS. We used Oprofile [19] to capture L2 and L3 cache misses, instruction counts, accesses to memory, and CPU utilization data. We collected this data every second, sending it to a central node for display. We do not claim that our approach rivals the sophistication of Hadoop (i.e., production code). Instead, our goal was to understand the impact of adding monitoring on BTS. We observed that BTS with monitoring suffered a 21% slowdown on MB-sized jobs, due to the increased startup overhead. On GB-sized jobs or larger, the runtime overhead caused an additional 15% slowdown. Despite these delays, BTS with monitoring still speeds up JLH by 2.5X on small jobs and 1.5X on larger jobs.

EAGLET allows scientists to test genetic hypotheses before sending them away for costly lab work. This process could proceed much faster if it were interactive. Before this work, we observed that vanilla EAGLET (i.e., without Hadoop or BashReduce) took an hour to complete a 230 MB job on a type 2 node; it was not designed for parallel execution. Running EAGLET within Hadoop and BashReduce platforms improved performance by using all available cores. Figure 11 shows BTS's speedup over VH.

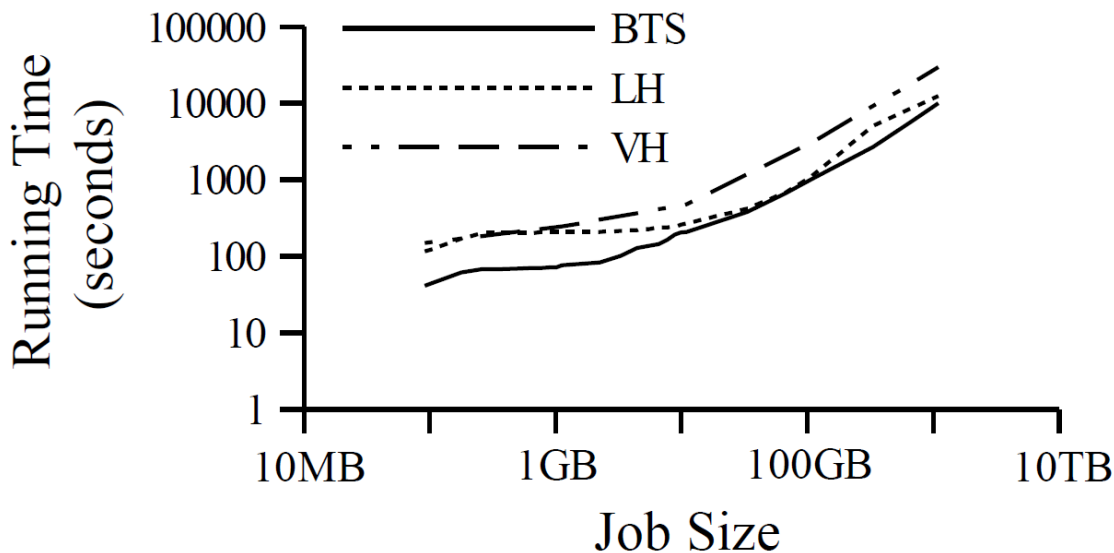

Figure 11: Comparison of BTS to VH and LH in terms of running time. Note log-log scale

These tests used 72 cores of type 2 hardware. With BTS, we completed a 91 MB job in 40 seconds. The same job took 150 seconds to run on VH. A 230 MB job ran on BTS in 68 seconds, a 59X speedup over vanilla EAGLET on 12 cores. For comparison to the state of the art, recent studies with CloudBlast, a competing tool for secondary genetic analysis, achieved 60 Mb/s [30] and 24 Mb/s [31]. BTS sustains 117 Mb/s. Note, these results are anecdotal. We cannot compare them directly because the workloads differ.

We also compared against LH. LH suffered from high startup costs when job sizes were small, essentially matching VH up to 1.1 GB sized jobs. It never achieved response times within 100 seconds. As job size increased, LH approached BTS performance. However, BTS (due to task scheduling) maintained 25% throughput gain even under a 1 TB job size.

4.2.3 Elasticity

Figure 12 shows throughput as we changed the number of cores in BTS. The platform scaled linearly up to 1 TB job. These tests were conducted on a 1 Gb/S network. The 72-core test (i.e., 6 type 2 nodes) produced results at 45% of network capacity. In Figure 12, regions where 72-core throughput equalled 36-core performance reflected startup costs. Large job sizes amortize these costs. For interactive workloads that run small jobs, however, the 72-core tests wasted resources. Managers should scale out until additional cores provide diminishing returns and no further.

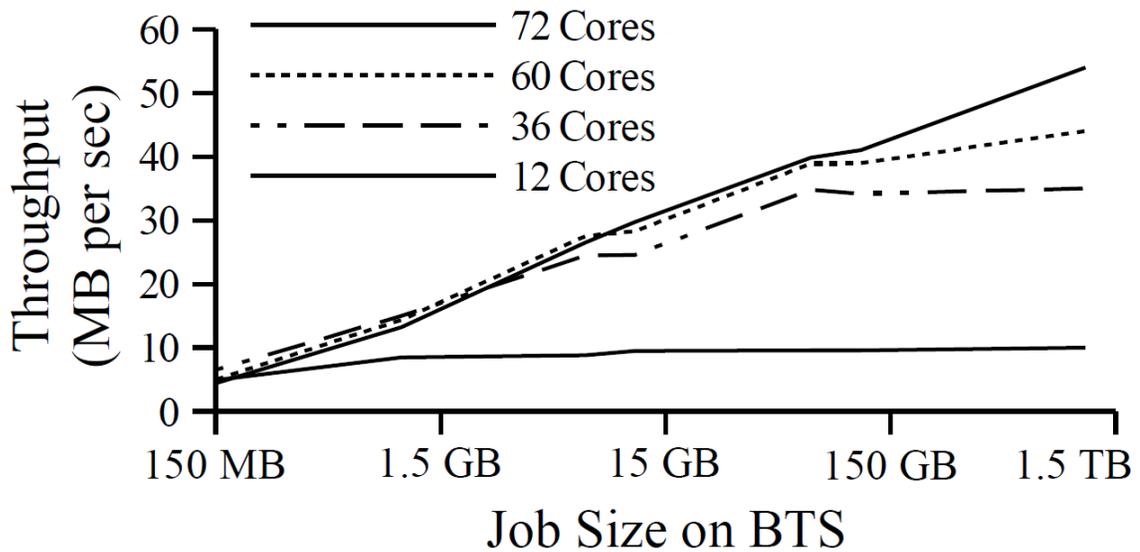

Figure 12: EAGLET on BTS as number of cores changed

Service-level objectives guarantee that a job will finish within a fixed running time [6], [32], [41], [42]. For data processing workloads, a job's running time depends on its size and the platform's achieved throughput at that size. If the job size is too small, startup costs dominate, limiting the data that can be processed within the fixed running time. Figure 13 shows BTS performance under various service level objectives. Each result reflects the platform configuration with highest achieved throughput within the fixed running time. Note, the 72-core case was only the best for 2-minute and 5-minute bounds. It has high startup costs, which allows the 36-core and 12-core case to perform better under tight bounds. Figure 13 shows performance relative to BTS's peak throughput without any service level objective. For reference, we also show the fixed running time relative to the running time when peak throughput was achieved. We observed that under a 2 minute

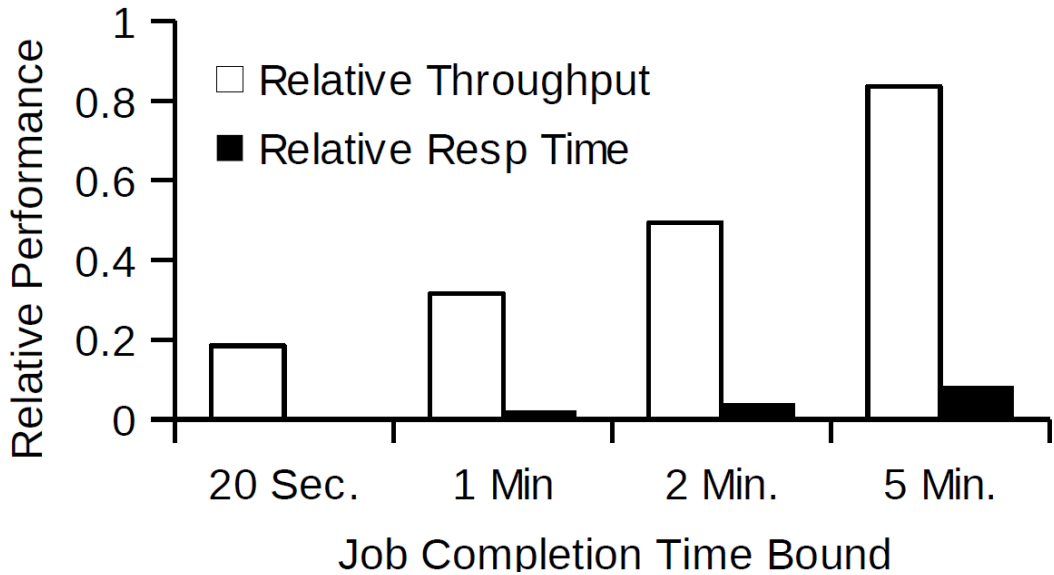

Figure 13: The throughput and running time of EAGLET on BTS clusters scaled to efficiently meet service level objectives

SLO BTS achieved 50% of its peak throughput. For reference, a 2 minute SLO represents 4% of the 50 minute run time needed to achieve peak throughput on 72 cores. A 5 minute SLO achieved 83% of peak throughput.

4.2.4 Virtualization and Heterogeneity

We tested our workloads on user-mode Linux virtual machines. For these tests, we used the original datasets for each workload. Each virtual machine was allocated 1 AMD Opteron core (i.e., type 3 in Table 2). We re-ran our task sizing algorithm on this hardware;

EAGLET had a kneepoint at a task size of 1.2 MB and Netflix had a kneepoint at a task size of 1 MB. Compared to type 2 hardware, i.e., without virtualization, we observed slowdown of 16% across both workloads. BTS still scaled out well, Figure 14 shows linear improvement for the Netflix workload.

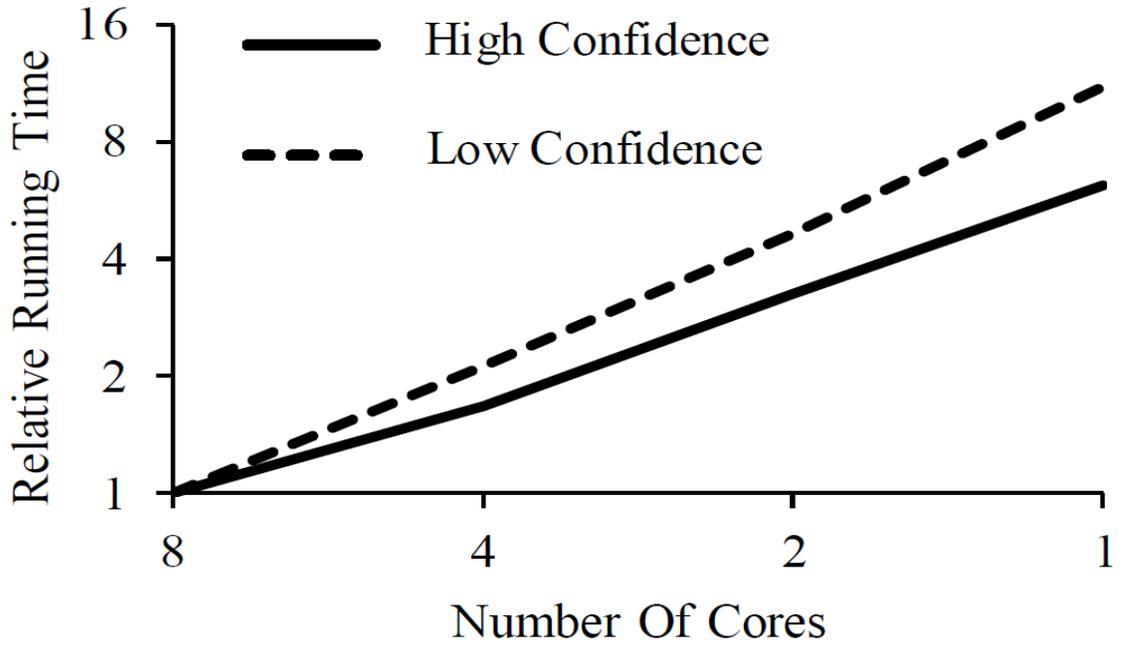

Figure 14: Netflix workload as cores scale on Type 3

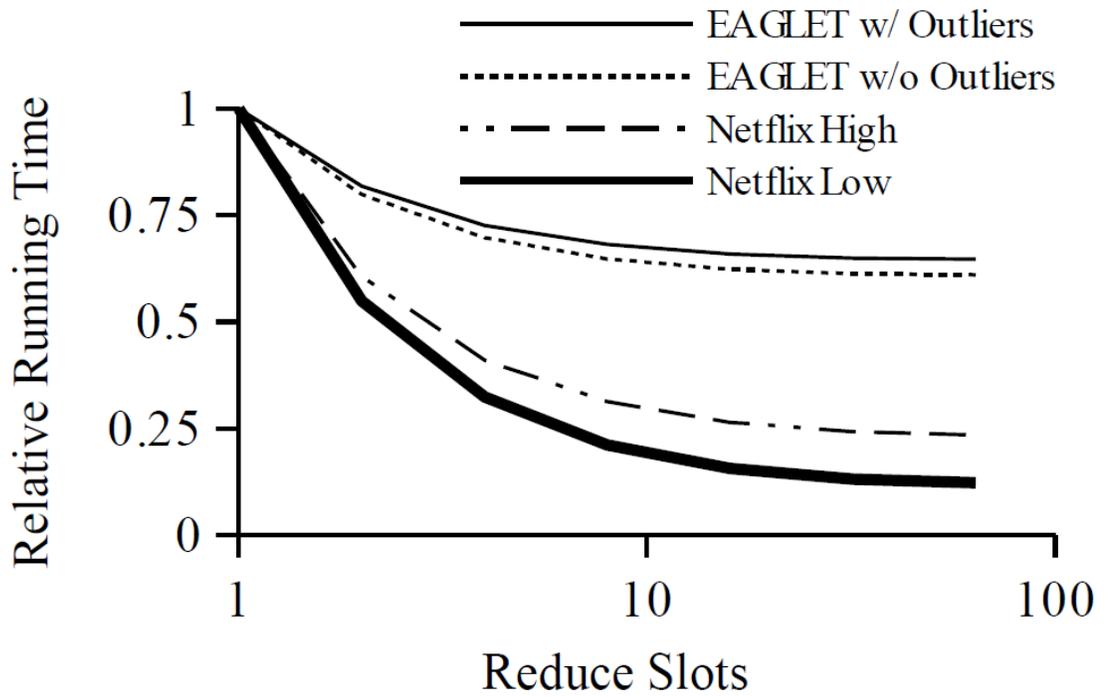

Figure 15: Netflix workload as job size increases

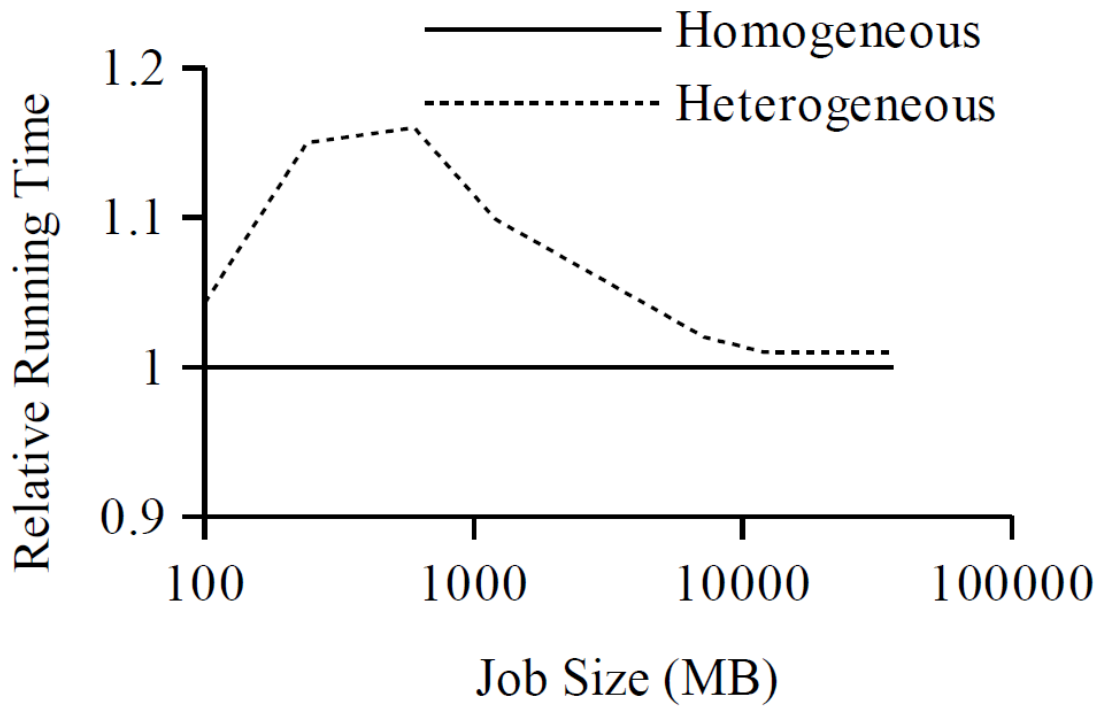

Figure 16: Network resource demand increases

We tested BTS under a heterogeneous environments where 12 of 60 cores were 15% slower than the others (i.e., 1 slow node). The slow node was of type 1 hardware and the others were of type 3 hardware. The slow nodes caused proportional slowdown on MB-sized jobs. However, as job size grew, BTS's round robin scheduler skipped over busy, slower cores, assigning more tasks to the faster cores. As a result, the performance loss is divided across 48 cores.

Finally, we studied the impact of reduce tasks. The BashReduce platform does not support multiple reduce slots gracefully. It requires mapping data back to all nodes and running the reduce stage as a map stage in an interactive computation. We used simulation to understand the impact of multiple reduce stages, and corresponding communication delay. We used formulas from [41] to understand the expected performance as reduce tasks increase. We calibrated these models with average map time, reduce time, and shuffle time from our experiments with 1-node map reduce. Figure 16 highlights the results. With EAGLET, secondary genetic analysis is compute intensive [30]. As a result, adding reduce tasks quickly exhibits diminishing returns. The Netflix workload, however, can speed up at the reduce stage.

# Chapter 5: Conclusion

Many workloads now consist of more data than data-parallel platforms can process within interactive response time constraints. Subsampling reduces processing requirements while providing statistical confidence on the accuracy of results.

In this thesis, we studied subsampling workloads, showing that subsampling from a large working set can significantly degrade cache locality. We made a case for tiny tasks, i.e., splitting subsampling workloads into many tasks with small working sets. Tiny tasks offer improved cache locality but suffer from scheduling overheads. We contend that scheduling overheads can be managed. Subsampling workloads benefit from task sizing to reduce cache miss rates and runtime costs. We implemented an algorithm to size tasks at sharp increases in cache miss rate within the BashReduce scheduler to reduce runtime overheads. We implemented a dynamic scheduler and data distribution layer in BashReduce. We validated our improved BashReduce against existing data-parallel platforms across multiple workloads. First, different platforms exhibit very different scheduling overheads depending on their objectives. Platforms designed for task-level recovery have overheads that are too high for tiny tasks. Platforms designed for job-level recovery perform better. Second, we show that task sizing can amortize some scheduling overheads with only a small increase in cache miss rate. Third, we show that our scheduling and data distribution platform on top of BashReduce increased throughput of the system. Our approach uses kneepoints on the task size to miss rate curve to determine task size. We demonstrated the benefit of our approach using genetic analysis and e-commerce datasets. In short, interactive workloads, our improved platform performed 9X better than vanilla Hadoop.

# References

[1] Sundeep Kambhampati, Jaimie Kelley, William C.L. Stewart, Christopher Stewart, and Rajiv Ramnath , Managing Tiny Tasks for Data-Parallel, Subsampling Workloads. In IEEE International Conference on Cloud Engineering Boston, MA, 2014

[2] F. Ahmad, S. Chakradhat, A. Raghunathan, and T. Vijaykumar. Tarazu: Optimizing MapReduce on Heterogeneous Clusters. In ACM ASPLOS, 2012.

[3] G. Ananthanarayanan, S. Kandula, A. Greenberg, I. Stoica, Y. Lu, B. Saha, and E. Harris. Reining in the outliers in map-reduce clusters using mantri. In USENIX Symp. on Operating Systems Design and Implementation, 2010.

[4] J. Badner and E. Gershon. Meta-analysis of whole-genome linkage scans of bipolar disorder and schizophrenia. Molecular psychiatry, 7(4):405–411, 2002.

[5] H. Bjornsson, G. Chockler, T. Saemundsson, and Y. Vigfusson. Dynamic performance profiling of cloud caches. In Tech Report, 2013.

[6] S. Bouchenak. Automated control for sla-aware elastic clouds. In Workshop on Feedback Computing, 2010.

[7] L. Bradshaw. Big data and what it means. U.S. Chamber of Commerce Foundation, 2013.

[8] G. Chockler, G. Laden, and Y. Vigfusson. Design and implementation of caching ervices in the cloud. In IBM Technical Report, 2012.

[9] P. Costa. Bridging the gap between applications and networks in data centers. In Sigops, 2013.

[10] R. Crowley. BashReduce - Crowley Code. https://github.com/erikfrey/bashreduce.

[11] N. Deng, C. Stewart, J. Kelley, D. Gmach, and M. Arlitt. Adaptive green hosting. In International Conference on Autonomic Computing, 2012.

[12] C. Ding and Y. Zhong. Predicting whole-program locality through reuse distance analysis. In ACM SIGPLAN Notices, volume 38, 2003.

[13] D. Ford, F. Labelle, F. I. Popovici, M. Stokely, V. Truong, L. Barroso, C. Grimes, , and S. Quinlan. Availability in globally distributed storage systems. In USENIX Symp. on Operating Systems Design and Implementation, 2010.

[14] I. n. Goiri, R. Beauchea, K. Le, T. D. Nguyen, M. E. Haque, J. Guitart, J. Torres, and R. Bianchini. Greenslot: scheduling energy consumption in green datacenters. In Proceedings of 2011 International Conference for High Performance Computing, Networking, Storage and Analysis, 2011.

[15] I. n. Goiri, K. Le, T. D. Nguyen, J. Guitart, J. Torres, and R. Bianchini. Greenhadoop: leveraging green energy in data-processing frameworks. In European Conference on Computer Systems, 2012.

[16] J. Kelley and C. Stewart. Balanced and predictable networked storage. In International Workshop on Data Center Performance, 2013.

[17] J. Kelley, C. Stewart, Y. He, and S. Elnikety. Cache provisioning for interactive NLP services. In Workshop on Large and Distributed Systems (LADIS), 2013.

[18] Z. Khayyat, K. Awara, A. Alonazi, H. Jamjoom, D. Williams, and P. Kalnis. Mizan: a system for dynamic load balancing in large-scale graph processing. In European Conference on Computer Systems, 2013.

[19] J. Levon and P. Elie. OProfile - A System Profiler for Linux. http://oprofile.sourceforge.net/.

[20] T. Luo, R. Lee, M. Mesnier, F. Chen, and X. Zhang. hstorage-db: heterogeneity-aware data management to exploit full capacity of hybrid storage systems. In VLDB, 2012.

[21] F. McSherry, R. Isaacs, M. Isard, and D. Murray. Differential dataflow. In CIDR, 2013.

[22] S. Melnik, A. Gubarev, J. J. Long, G. Romer, S. Shivakumar, M. Tolton, and T. Vassilakis. Dremel: Interactive analysis of web-scale datasets. In Proc. of the 36th Int'l Conf on Very Large Data Bases, 2010.

[23] M. Mitzenmacher. The power of two choices in randomized load balancing. IEEE Transactions on Parallel and Distributed Systems, 2001.

[24] A. Nandi, C. Yu, P. Bohannon, and R. Ramakrishnan. Distributed cube materialization on holistic measures. 2011.

[25] National Human Genome Research Institute. Dna sequencing costs.

http://www.genome.gov/sequencingcosts/, 2013.

[26] K. Ousterhout, A. Panda, J. Rosen, S. Venkataraman, R. Xin, S. Ratnasamy, S. Shenker, and I. Stoica. The case for tiny tasks in compute clusters. In HotOS, 2013.

[27] K. Ousterhout, P. Wendell, M. Zaharia, and I. Stoica. Sparrow: Scalable scheduling for sub-second parallel jobs. In ACM Symp. on Operating Systems Principles, 2013.

[28] D. A. Patterson and J. L. Hennessy. Computer Organization and Design, Fourth Edition, Fourth Edition: The Hardware/Software Interface (The Morgan Kaufmann Series in Computer Architecture and Design). Morgan Kaufmann Publishers Inc., San Francisco, CA, USA, 2008.

[29] R. Power and J. Li. Piccolo: Building fast, distributed programs with partitioned tables. In USENIX Symp. on Operating Systems Design and Implementation, 2010.

[30] A. Rasmussen, M. Conley, R. Kapoor, V. Lam, G. Porter, and A. Vahdat. Themis: An i/o efficient mapreduce. In ACM Symp. on Cloud Computing, Oct. 2012.

[31] M. Schatz. Cloudburst: highly sensitive read mapping with mapreduce. In BioInformatics, 2009.

[32] C. Stewart, A. Chakrabarti, and R. Griffith. Zoolander: Efficiently meeting very strict, low-latency slos. In International Conference on Autonomic Computing, 2013.

[33] C. Stewart, K. Shen, A. Iyengar, and J. Yin. Entomomodel: Understanding and avoiding performance anomaly manifestations. In IEEE International Symposium on Modeling, Analysis, and Simulation of Computer and Telecommunication Systems, 2010.

[34] W. C. Stewart, E. N. Drill, D. A. Greenberg, et al. Finding disease genes: a fast and flexible approach for analyzing high-throughput data. European Journal of Human Genetics, 19(10):1090, 2011.

[35] C. Tsai, J. Chou, and Y. Chung. Value-based tiering management on heterogeneous block-level storage system. In CloudCom, 2012.

[36] Y. Wang, J. Tan, W. Yu, L. Zhang, and X. Meng. Preemptive reducetask scheduling for fair and fast job completion. In Int'l Conf. on Autonomic Computing.

[37] A. Waterland, J. Appavoo, and M. Seltzer. Parallelization by simulated tunneling. In Workshop on Hot Topics in Parallelism, 2012.

[38] T. White. Hadoop: The definitive guide. O'Reilly Media.

[39] J. Xie. Improving mapreduce performance in heterogeneous Hadoop clusters. In Intn'l Heterogeneity in Computing Workshop, 2010.

[40] M. Zaharia, M. Chowdhury, T. Das, A. Dave, J. Ma, M. McCauley, M. J. Franklin, S. Shenker, and I. Stoica. Resilient distributed datasets: A fault-tolerant abstraction for in-memory cluster computing. In USENIX Symp. on Networked Systems Design and Implementation, 2012.

[41] Z. Zhang, L. Cherkasova, and B. Loo. Performance modeling of mapreduce jobs in heterogeneous cloud environments. In IEEE CLOUD, 2013.

[42] Z. Zhang, L. Cherkasova, A. Verma, and B. Loo. Automated profiling and resource management of pig programs for meeting service level objectives. In Int'l Conf. on Autonomic Computing, 2012.

[43] Netflix prize. http://www.netflixprize.com//index.

[44] Apache Cassandra http://cassandra.apache.org/

[45] Hong Xu, Baochun Li, RepFlow: Minimizing Flow Completion Times with Replicated Flows in Data Centers. Proc. IEEE Conference on Computer Communications (IEEE INFOCOM), 2014.

[46] Yanfei Guo, Jia Rao, and Xiaobo Zhou, iShuffle: Improving Hadoop Performance with Shuffle-on-Write. In International Conference on Autonomic Computing, 2013

[47] Niranjan Kamat, Prasanth Jayachandran, Kathik Tunga, Arnab Nandi, Distributed Interactive Cube Exploration. In IEEE International Conference on Data Engineering, 2014.

[48] Derek G. Murray, Frank McSherry, Rebecca Isaacs, Michael Isard, Paul Barham, and Martín Abadi. 2013. Naiad: a timely dataflow system. In ACM Symposium on Operating Systems Principles (SOSP), 2013.